# A Novel Injection-Locked *S*-Band Oven Magnetron System Without Waveguide Isolators

Shaoyue Wang, *Student Member, IEEE*, Yang Shen, Chongwei Liao, Jianwei Jing, and Changjun Liu, *Senior Member, IEEE*

*Abstract*— A novel injection-locked *S*-band microwave oven magnetron system is proposed, analyzed, and experimented. The magnetron is considered as a two-port oscillator, and the filament structures of it are used as a port for injection locking. To our knowledge, it is the first time that microwaves have been injected through magnetron filaments. The microwave is injected into the magnetron filter box using a self-designed injection structure, entering the resonant cavity through filament leads. The intrinsic isolation between the magnetrons' output and filament is utilized as an isolator. No waveguide circulators or couplers between the injection source and the magnetron were used in the system, resulting in low cost and compact injection locking. Two types of injection structures were performed. The maximum locking bandwidth of 0.7 MHz was achieved at an injection ratio of 0.2. The weight and volume of the proposed injection-locked magnetron system were reduced to 28% and 16%, respectively. The magnetron's output is successfully locked by an external signal. The experiment results reveal that the proposed low-cost injection-locked oven magnetron system works reliably. It has significant potential applications for future microwave ovens with frequency-selective heating to improve both efficiency and uniformity.

*Index Terms*— Compact, filament, injection locking, low cost, magnetron.

## I. INTRODUCTION

**M**AGNETRONS have occupied an important position in industrial microwave applications recently because of their low cost, high efficiency, and compact volume. Magnetrons have been widely used in microwave industries, especially in microwave heating applications [1], [2], [3], [4], [5]. The phase and frequency of a magnetron are unstable, and its output characteristic is easily affected by external factors, such as the applied voltage and temperature [6], [7], [8]. In areas requiring the distribution and intensity of microwave field in the heating cavity to be controllable or high precision, a free-running magnetron cannot be applied directly because of its random frequency and phase. Furthermore, its noisy spectrum easily affects other electronic devices with similar operating frequencies [9], [10], [11], [12].

Injection-locking technology, which has been studied for many years, is an important method to improve the quality of a magnetron's output and control its phase and amplitude. Adler [13] studied the theory of oscillator injection-locking and obtained the famous Adler equation. Based on the equivalent circuit theory, Slater [14] and David [15] theoretically analyzed the effects of frequency pulling and frequency pushing of injection-locked magnetrons.

Many additional investigations on injection-locked magnetron systems have been conducted, and more applications of injection-locked magnetrons have been identified. Tahir et al. [16], [17] conducted injection-locking experiments on a 1-kW microwave oven magnetron with a phase-locked loop to analyze its noise and locking performance and expanded the injection locking to digital modulation. Liu et al. [18] achieved a four-way 2.45-GHz continuous-wave (CW) magnetron power-combining system—a total output power over 60.6 kW was achieved with a power-combining efficiency greater than 91.5%. Yang et al. [19] built a simultaneous wireless information and power transfer system based on a phased array using four 5.8-GHz injection-locked magnetrons. Yang et al. [20] proposed a novel magnetron microwave sweep frequency heating method based on injection-locked magnetrons, and the coefficient of variation of the potato heated was reduced from 0.699 to 0.535, which presents a successful application of injection-locking technique. Chen et al. [21] proposed a power-combining system based on two 5.8-GHz magnetrons with a waveguide magic tee instead of ferrite circulators, efficiently reducing the insertion loss and achieving a combining efficiency of 94.7%. These studies illustrate the potential applications of injection-locked magnetrons. However, ferrite circulators or waveguide devices that isolate the injection signal from the magnetron output are usually large and costly, and limit the application of injection-locked magnetrons in heating industry and microwave oven.

In this article, we propose a novel injection-locked *S*-band 1-kW cook-type magnetron system using a method without any waveguide circulators or other waveguide devices to isolate the injection signal from the magnetron's output. The method greatly reduces the cost of the injection-locked magnetron system and is expected to be applied to industrial microwave heating and microwave ovens to achieve

Manuscript received 8 November 2022; revised 31 December 2022; accepted 28 January 2023. Date of publication 16 February 2023; date of current version 24 March 2023. This work was supported in part by the Natural Science Foundation of China under Grant U22A2015 and Grant 62071316 and in part by the Sichuan Science and Technology Program under Grant 2021YFH0152. The review of this article was arranged by Editor S. Prasad. *(Corresponding author: Changjun Liu.)*

The authors are with the School of Electronics and Information Engineering, Sichuan University, Chengdu 610064, China, and also with the Yibin Industrial Technology Research Institute, Sichuan University, Yibin 644000, China (e-mail: cjliu@scu.edu.cn).

Color versions of one or more figures in this article are available at https://doi.org/10.1109/TED.2023.3243041.

Digital Object Identifier 10.1109/TED.2023.3243041

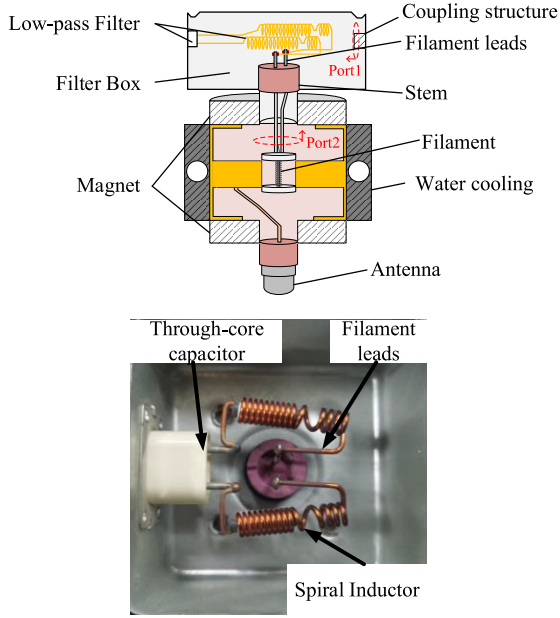

Fig. 1. Diagram of magnetron and photograph of filter box.

controllable heating with low cost. The magnetron's filter box was reformed to provide an injection path of the external signal. The injection-locked magnetron system based on two different proposed injection structures is implemented in this article, with a maximum locking bandwidth of 0.7 MHz at an injection power ratio of 0.2. The frequency of magnetrons in the two systems is successfully injected at the frequency of the injected signal, frequency-locked magnetron in the locking range. Frequency-locked magnetron is realized in the injection-locking system. The differences between the novel system and the conventional one are also presented. This article presents a novel idea for the research of injection-locked magnetrons.

## II. THEORETICAL ANALYSIS

### A. Analysis of Magnetrons

The structure of an S-band 1-kW cook-type magnetron,[1] including its filter box, is shown in Fig. 1. A magnetron is always considered as a one-port oscillator composed of a resonant cavity, an output antenna, a magnetic subsystem, a filter box, and a water-cooling component, and the microwave power is mainly output from the antenna [22]. However, microwave leakage from a magnetron is inevitable. There is about one-thousandth of the output microwave power of an operating magnetron being coupled to the filter box through filament leads from the magnetron's resonant cavity. The filament leads and the filter box is equivalent to a port. Thus, a magnetron in this article is described as a two-port oscillator and the equivalent circuit of it is shown in Fig. 2(a). As it can be seen, the equivalent circuit includes the equivalent electronic admittance $g + jb$, the equivalent oscillating circuit with parameters $L$, $C$, and $R$, the equivalent filament admittance $G_1 + jB_1$, and the equivalent load admittance $G_2 + jB_2$. $I_1$, $V_{RF0}$, and $I_{RF0}$ are the amplitudes of the current of the leakage power from the filament, the voltage, and the current of the magnetron's output, respectively.

[1]"Magnetron" in this paper refers to an S-band oven magnetron.

When the magnetron satisfies the single-mode oscillation condition, the circuit equation can be written as

$$g + jb = \frac{1}{R} + j\omega_c C + \frac{1}{j\omega_c L} + G_1 + jB_1 + G_2 + jB_2 \quad (1)$$

where $\omega_c$ is the free-running frequency of the magnetron, and the equivalent filament admittance is expressed as

$$G_1 + jB_1 = \frac{\widetilde{I}_1}{\widetilde{V}_{RF0}}. \quad (2)$$

A novel injection-locked magnetron based on "filament-injection" method is proposed from this two-port model. We assume that a reference signal at frequency $\omega_{inj} \approx \omega_c$ is injected through the filament leads. Then, the updated equivalent circuit is shown in Fig. 2(b). The voltage and current of the injection source are $V_{inj0}$ and $I_{inj0}$, respectively. The voltage and current entering the magnetron are represented by $I_{inj} = kI_{inj0}$ and $V_{inj} = kV_{inj0}$, $0 < k < 1$, respectively, since there is an attenuation in the injection path between the magnetron resonant cavity and the injection source. Therefore, the updated equivalent filament admittance with the assumption of $V_{inj0} \ll V_{RF0}$ is

$$G'_1 + jB'_1 = \frac{\widetilde{I}_{inj} + \widetilde{I}_1}{\widetilde{V}_{inj} + \widetilde{V}_{RF0}} = \frac{I_1}{V_{RF0}} \left( \frac{1 + \frac{I_{inj}}{I_1} e^{j(\omega_{inj}-\omega'_c)t}}{1 + \frac{V_{inj}}{V_{RF0}} e^{j(\omega_{inj}-\omega'_c)t}} \right)$$
$$\approx G_1[1 - (m+1)k\rho e^{j\theta}] \quad (3)$$

where $\theta = (\omega_{inj} - \omega'_c)t$ is the phase difference between the magnetron's output and injected signal, $\omega'_c$ is the output frequency of the magnetron under the influence of the injected signal, $m = I_{RF0}/I_1 \gg 1$ is a constant related to the structure and load of the magnetron, $G_1 = I_1/V_{RF0}$, and the injection ration $\rho$ is defined as

$$\rho = \frac{I_{inj0}}{I_{RF0}} = \frac{I_{inj}}{mkI_1}. \quad (4)$$

According to Slater's work [14], the equivalent electronic admittance of the magnetron is given as follows:

$$g = \frac{1}{R}\left(\frac{V_{dc}}{V_{RF}} - 1\right) \quad (5)$$
$$b = b_0 + g \tan\alpha \quad (6)$$

where $V_{dc}$ is the voltage between the anode and cathode of magnetrons; $V_{RF}$ is the amplitude of the voltage of the magnetron actual output; $b_0$ is a constant; and $\alpha$ is defined as the pushing parameter, its unit is rad, and its value varies with the structure of magnetrons.

Then, because of the assumption of $\omega_{inj} \approx \omega_c$, the equivalent admittance of resonant cavity is obtained

$$j\omega'_c C + \frac{1}{j\omega'_c L} \approx 2jC(\omega'_c - \omega_0) \quad (7)$$

where $\omega_0$ is the resonant frequency of the magnetron's resonant cavity. Substituting (3), (6), and (7) into (1) and using the Euler formula, we then obtain

$$g + jb = \frac{1}{R} + 2jC(\omega'_c - \omega_0) + G_1$$
$$[1 - (m+1)k\rho(\cos\theta + j\sin\theta)] + G_2 + jB_2. \quad (8)$$

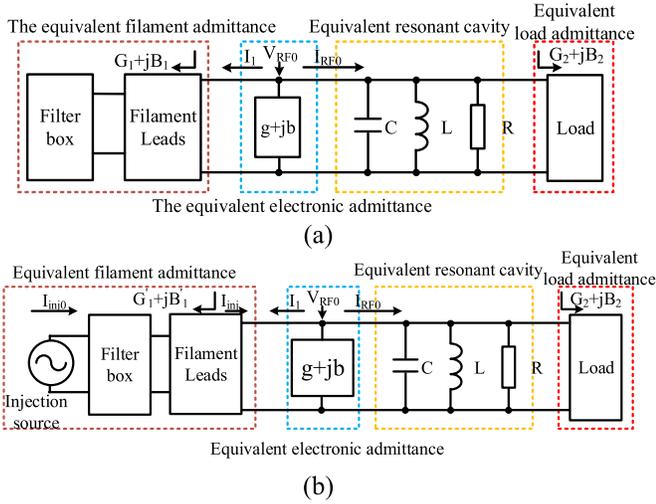

Fig. 2. Equivalent circuit of magnetrons. (a) Equivalent circuit of a free-running magnetron. (b) Equivalent circuit of an injection-locked magnetron.

Referring to the equivalent circuit theory, we have $Q_L = \omega_0 C/g$, $Q_f = \omega_0 C/G_1$, $Q_a = \omega_0 C/G_2$. $Q_L$, $Q_f$, and $Q_a$ are the loaded quality factor, the external quality factor of the filament port, and the output antenna port of the magnetron. Also, we derived the imaginary part of (8), which can be simplified as

$$\omega_c' = \omega_0 + \frac{b_0 - B_2}{2C} + \frac{\tan\alpha}{2Q_L} - \frac{(m+1)k\omega_0}{2Q_f}\rho\sin\theta. \quad (9)$$

The first three terms on the right-hand side of (9) represent the free-running frequency without injected signal and will denote it with $\omega_c$ below. In (9), the denominator of the fourth term is the right-hand side, and $Q_f$ has a mathematical relation with $Q_a$. It has been known from the derivation above that $G_1 = I_1/V_{RF0} = I_{RF0}/mV_{RF0} = G_2/m$, and $Q_f = \omega_0 C/G_1 = m\omega_0 C/G_2 = mQ_a$. Then, we subtract $\omega_{inj}$ from both sides of (9) set $d\theta/dt = \omega_{inj} - \omega_c'$, and (9) can be simplified as

$$\frac{d\theta}{dt} = \omega_{inj} - \omega_c + \frac{k\omega_0}{2Q_{ext}}\rho\sin\theta \quad (10)$$

$$\frac{1}{Q_{ext}} = \frac{1}{Q_a} + \frac{1}{Q_f} \quad (11)$$

where $Q_{ext}$ in (10) and (11) is the external quality factor of the magnetron. Equation (10) describes the variation speed of the magnetron's output phase when an external signal is injected into it. For the steady state of the magnetron, $d\theta/dt$ must be zero so that the locking condition can be derived from (10)

$$\Delta\omega = \omega_{inj} - \omega_c = \frac{k\omega_0}{2Q_{ext}}\rho\sin\theta \quad (12)$$

or

$$|\Delta\omega| = |\omega_{inj} - \omega_c| \leq \frac{k\omega_0}{2Q_{ext}}\rho. \quad (13)$$

Apparently, (12) and (13) have almost the same form with the famous Adler equation [13], and when $k = 1$, they will be reduced to the result presented in [13]. We can know that the filament-injection method follows the identical law with the existing method. Also, the injection ratio $\rho$ is strongly related to the locking bandwidth: the higher the injection ratio, the broader the locking bandwidth. However, unlike the Adler equation, there is a coefficient $k$ whose range of value is 0–1 in (12) and (13). The coefficient indicates that not all the power injected can enter the magnetron's cavity. The value of $k$ is an important factor that restrains the locking bandwidth of the proposed method. Therefore, in order to guide the design of the injection structure, it is necessary to find out the parameters that directly or indirectly affect $k$.

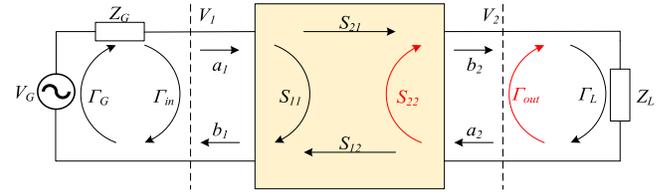

Fig. 3. Diagram of the two-port network.

In addition, it should be noted that the constant $m$, which is defined as the ratio of $I_{RF0}$ and $I_1$, shows that there is an intrinsic isolation between the magnetron's output and the filament leads or the injection source. The phenomenon indicates that the proposed method may eliminate the high-power isolation devices in an injection-locked magnetron system, such as circulators, and will greatly reduce the cost and volume of the system.

### B. Analysis of the Injection Structure

Fig. 1 shows that the filter box of a magnetron contains filament leads, helical coil inductors with ferrite cores, and two feed-through capacitors. The inductors and capacitors form low-pass filters that depress the leakage of microwave power from the filament leads. The filter box will provide the necessary space for an injection-locking structure used in the proposed filament-injection method. Thus, combined with the analysis in Section II-A, the filament-injection method can be performed through the filter box.

In this case, we assume that there are two virtual power coupling ports in the filter box to explore the design parameters of the structure and their relationship with $k$. A microwave source is applied to generate a signal and inject it into the magnetron at port 1. Thus, we regard it as a two-port network; Ports 1 and 2 are shown in Fig. 1(a).

Fig. 3 shows the network view. $a_i$ and $b_i$ are the incident and reflected power waves of port $i$, and they satisfy

$$\tilde{a}_i = \frac{\tilde{V}_i + Z_i\tilde{I}_i}{2\sqrt{Z_i}}, \tilde{b}_i = \frac{\tilde{V}_i - Z_i\tilde{I}_i}{2\sqrt{Z_i}} \quad (14)$$

$$\begin{bmatrix}\tilde{b}_1\\\tilde{b}_2\end{bmatrix} = \begin{bmatrix}S_{11} & S_{12}\\S_{21} & S_{22}\end{bmatrix}\begin{bmatrix}\tilde{a}_1\\\tilde{a}_2\end{bmatrix} \quad (15)$$

where $i$ represents the port number, $Z_i$ is the system impedance of port $i$, and $V_i$ and $I_i$ are the voltage and current of port $i$, respectively. Also, the output power $P_{inj}$ and the input power $P_{inj0}$ of the network are

$$P_{inj0} = \frac{1}{2}|a_1|^2, \quad P_{inj} = \frac{1}{2}(|b_2|^2 - |a_2|^2). \quad (16)$$

Thus, after combining (14)–(16), the transmission efficiency (defined as the ratio of output power $P_{inj}$ to the input power

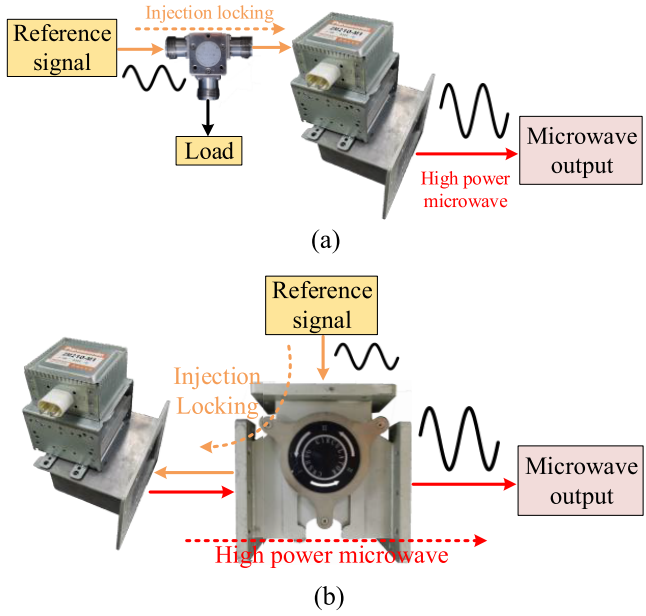

Fig. 4. Diagram of different injection-locked magnetron systems. (a) Filament and (b) conventional injection-locked magnetron system.

$P_{\text{inj0}}$) is deduced as

$$k^2 = \frac{P_{\text{inj}}}{P_{\text{inj0}}} = \frac{|S_{21}|^2(1-|\Gamma_L|^2)}{|1-S_{22}\Gamma_L|^2} \quad (17)$$

where $\Gamma_L$ is the load and input voltage reflection coefficients. There are many factors that affect the value of $k$, and the relationships between them are complex. $|S_{21}|$, $|S_{22}|$, and $\Gamma_L$ are all related to the magnetron structures, especially transmission structures. Substituting (17) into (13), we then obtain

$$|\Delta\omega| = |\omega_{\text{inj}} - \omega_c| \le \sqrt{\frac{|S_{21}|^2(1-|\Gamma_L|^2)}{|1-S_{22}\Gamma_L|^2} \frac{\omega_0}{2Q_{\text{ext}}}\rho}. \quad (18)$$

We can know from (18) that increasing the locking bandwidth $|\Delta\omega|$ requires improving $|S_{21}|$. Consequently, $|S_{21}|$ of the injection structure has the reference significance for its design.

### C. Injection-Locked Magnetron Systems

Injection locking refers to injecting an external low-power, high-quality signal into a large power oscillator (e.g., a magnetron). The injection-locked system is usually implemented with circulators. Waveguide circulators—often heavy and large—are used in high-power situations, such as an injection-locked magnetron system.

From the above analysis, we can know that the filament leads of a magnetron can also be considered as a port to inject an external signal. Moreover, the intrinsic isolation between the filament and high-power output can be instead of the high-power circulator to provide the injection path. The diagram of the injection-locked magnetron system with the proposed filament-injection method is shown in Fig. 4(a). For intuitive comparison, Fig. 4(b) gives out the diagram of a conventional injection-locked magnetron system. It can be seen that the proposed system eliminates the high-power, three-port waveguide circulator between the magnetron and the signal source. The only circulator used in the system is a low-power coaxial circulator to prevent the source from the power reflection caused by mismatch and unwanted leakage.

In some situations, four-port waveguide circulators and a high-power dummy load are used in a conventional system to protect the injecting system. Thus, the proposed filament-injection method will greatly reduce the cost and volume of an injection-locked magnetron system and will make the system widely applied in microwave heating or drying industry and even in microwave oven.

## III. NOVEL INJECTION-LOCKING STRUCTURES

According to the derivation and analysis in Section II, the proposed injection-locked magnetron system is feasible and low cost. The advantages of the system are obvious. However, the filament leads of a magnetron are always enclosed in the filter box, and there is no injection port, which can be used directly. It is important to design an effective injection structure. In this section, two different structures are proposed. Both two structures are inexpensive and easy to implement.

### A. Structure 1: Monopole Probe Injection Structure

The simulation model is shown in Fig. 5(a). A monopole probe is installed in the filter box. After a measurement test, the free-running frequency of the magnetron we used is around 2.43 GHz. Thus, the length of the monopole probe is approximately a quarter wavelength of 2.43-GHz microwave, and its ground is the metal shielding of the filter box. The reference signal is injected through the monopole probe and coupled with the spiral inductor and filament leads.

We used Ansys full-wave simulation software to analyze the filament structure of the magnetron and optimize the monopole probe position. The simulation results of the structure are shown in Fig. 5(b). The optimal length of the monopole is 24 mm. The results illustrate that $|S_{21}|$ is above −5 dB between 2.424 and 2.442 GHz. $|S_{21}|$ in the frequency range is not very good, but it is acceptable for microwave oven applications with the simple and low-cost injection structure. The results show that the proposed structure is efficient to couple microwave into the magnetron's resonant cavity. Therefore, the monopole probe in the filter box will be implemented according to the parameters achieved in the simulation.

### B. Structure 2: Near-Field Coupling Ring Structure

A near-field coupling structure can achieve high efficiency in short-distance power injection with proper parameters [23]. We introduced two near-field rings as a coupling structure for injection locking. As shown in Fig. 6, a receiving ring with a dc-blocking capacitor is connected to the filament leads, and the other transmitting ring is used as feeding. One ring lead is connected to the filter box metal shielding as ground, and another ring lead is connected to a coaxial adapter as the feeding port. Table I shows the detailed parameters of the proposed near-field coupling rings.

The simulation model is simplified that the copper rods are directly connected to the filament leads; the spiral inductance and through-core capacitance is equivalent to an open circuit at 2.45 GHz. The transmitting and receiving rings are concentric circles fabricated on the top and bottom sides of the printed circuit board (PCB) substrate (FR4 with a thickness of 1 mm). Because the receiving ring is welded on the filament leads, a gap has been designed in the middle of the receiving ring. A capacitor of 33 pf mounted on the gap provides a radio frequency (RF) path and blocks the filament voltage to avoid a short circuit. The injection signal is coupled from the transmitting ring to the receiving ring with this structure and finally fed into the magnetron's resonant cavity through the filament leads.

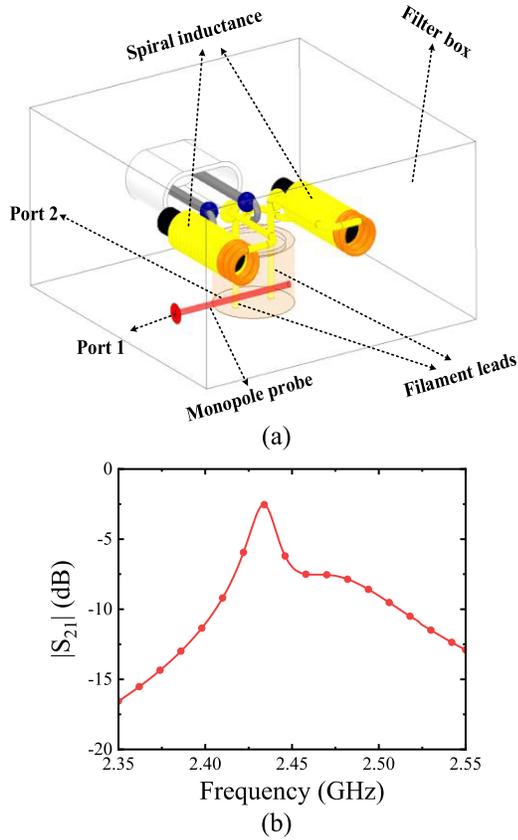

Fig. 5. Monopole antenna method. (a) Simulation model. (b) Simulation result.

TABLE I
DIMENSIONS OF THE RING S

| Symbol | Value |
|---|---|
| $R$ | 6.3 mm |
| $r$ | 4.5 mm |
| $W_1$ | 1 mm |
| $W_2$ | 1.5 mm |
| $W_3$ | 2 mm |
| $W_4$ | 4 mm |
| $d$ | 13 mm |
| $g_1$ | 2 mm |
| $g_2$ | 4 mm |
| $\theta$ | 5 deg |

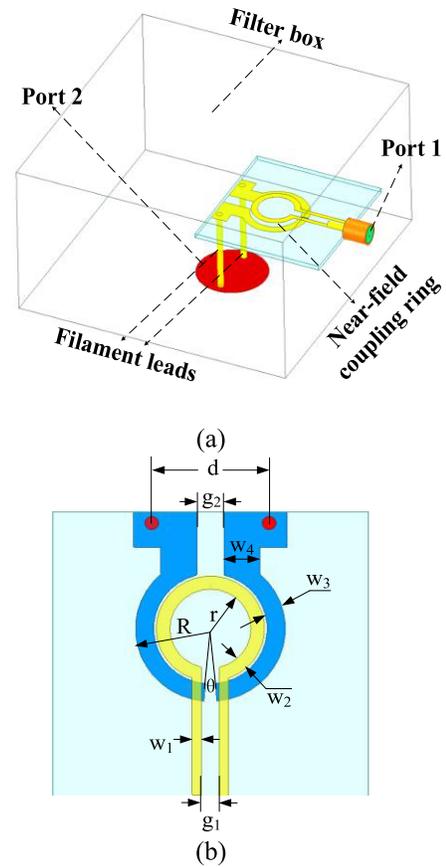

Fig. 6. Near-field ring coupling method. (a) Structures. (b) Coupling rings. (c) Simulated result.

The simulation results are shown in Fig. 6(c). The result demonstrates that $|S_{21}|$ is above −5 dB between 2.30 and 2.65 GHz, covering the operating frequency of the magnetron. Although $|S_{21}|$ is not ideal because there is some energy dissipating in the filter box and the substrate, the insertion loss is acceptable. The results prove that the structure can realize the desired function. Therefore, it is a suitable structure for coupling a reference signal into the magnetron.

## IV. EXPERIMENTS AND ANALYSIS

### A. Experiment Setup

In the experimental system, an S-band 1-kW magnetron (M1–2M210, Panasonic) was modified to introduce the injection locking through a filament. Fig. 7 shows the two injection-locking structures of the magnetron: a monopole probe and a coupling ring.

Fig. 7(a) shows a photograph of structure 1; the monopole probe position is 10 and 11.5 mm from the side and bottom of the magnetron filter box, respectively. Fig. 7(b) shows a photograph of structure 2, 4 mm from the top of the magnetron cover, and the received ring is connected to the filament leads.

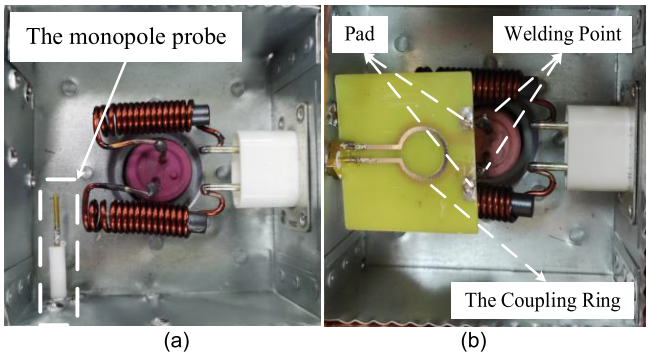

Fig. 7. Photographs of modified magnetron filter boxes. (a) Monopole probe. (b) Coupling ring.

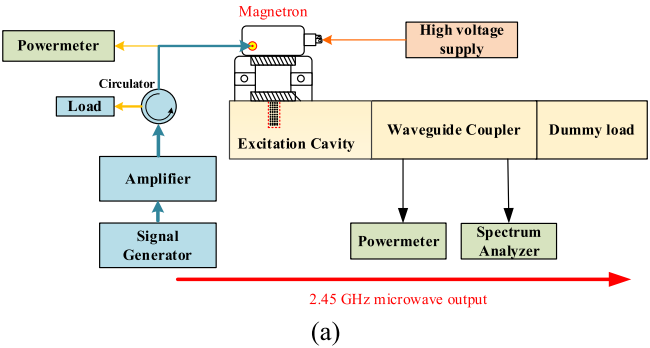

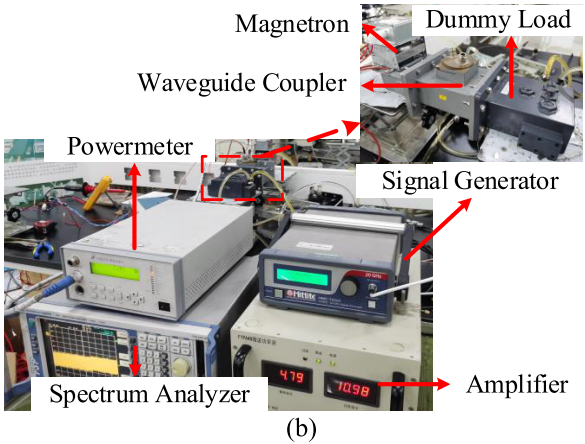

Fig. 8. Injection-locking experiment system. (a) Diagram and (b) picture of the system.

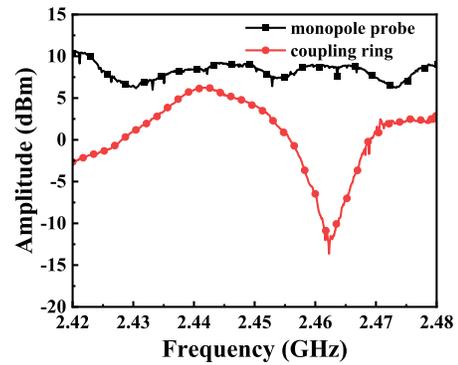

Fig. 9. Measurement output power of the cold test.

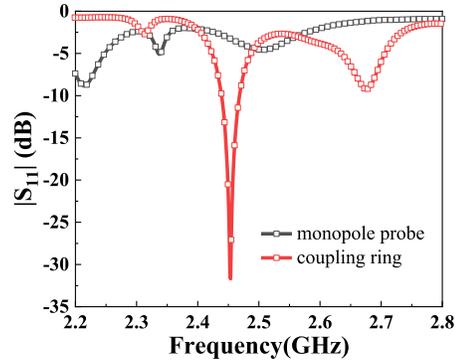

Fig. 10. Measurement $|S_{11}|$ of the two structures. Probe and a coupling ring.

function is connected to the output coupler. The spectrum analyzer is to detect the output power of the microwave injected by the monopole probe or the coupling ring from the waveguide. The power of the injected microwave is 35 dBm. The experiment is to replace the measurement of $|S_{21}|$ because $|S_{21}|$ cannot be measured in practice.

*Step 2:* The magnetron injection-locking experiments are conducted using the experimental system in Fig. 8. The external signal is injected through the proposed injection structures and observes the spectrum. Record the locking bandwidth under different injection ratios when the injection-locking phenomenon occurs.

The measurement system is shown in Fig. 8(a). The left side is the injection-locking system. A Hitter HMC-T2220 signal generator and a power amplifier are used to generate injecting microwaves up to 20 W. An AV2433 power meter is used to monitor the injecting power.

The modified magnetron is driven by a self-made low ripple dc power supply, which provides 4200-V anode voltage. The filament current is turned off after the magnetron operation stabilizes. A spectrum analyzer (FSP, R&S) and a power meter are used to observe the output spectrum and monitor the output power of the magnetron through a directional waveguide coupler. Fig. 8(b) shows the photographs of the system.

We will conduct the injection-locking experiments in the following two steps.

*Step 1:* A cold test is conducted on the designed injection structure. The frequency of the injected signal is tuned from 2.42 to 2.48 GHz, and a spectrum analyzer with the max-hold

### B. Results and Discussion

Fig. 9 shows the cold test results. The output power of the monopole probe and the coupling ring is above 6 dBm between 2.42 and 2.48 GHz, with a maximum of 10.6 dBm at 2.4204 GHz. Furthermore, the output power of the coupling ring is above 0 dBm between 2.428 and 2.455 GHz, with a maximum of 6.2 dBm at 2.4420 GHz. The results demonstrate that there is a signal path from the signal source to the waveguide, and the two proposed structures can successfully transmit the injected signal into the magnetron with an attenuation. The cold test $|S_{11}|$ results of the two structures are measured and shown in Fig. 10. $|S_{11}|$ is between –2.5 and –3.5 dB of the monopole probe structure with the frequency range of 2.42–2.45 GHz. and the frequency range when $|S_{11}|$ is below –10 dB of coupling ring structure is 2.32 and 2.34 GHz, and 2.44–2.46 GHz. However, the output power of the cold test shown in Fig. 9 and the locking bandwidth measured in Table I illustrate that the monopole probe has a better performance than the coupling ring. This phenomenon is due to the differences of coupling mechanism between the

1892

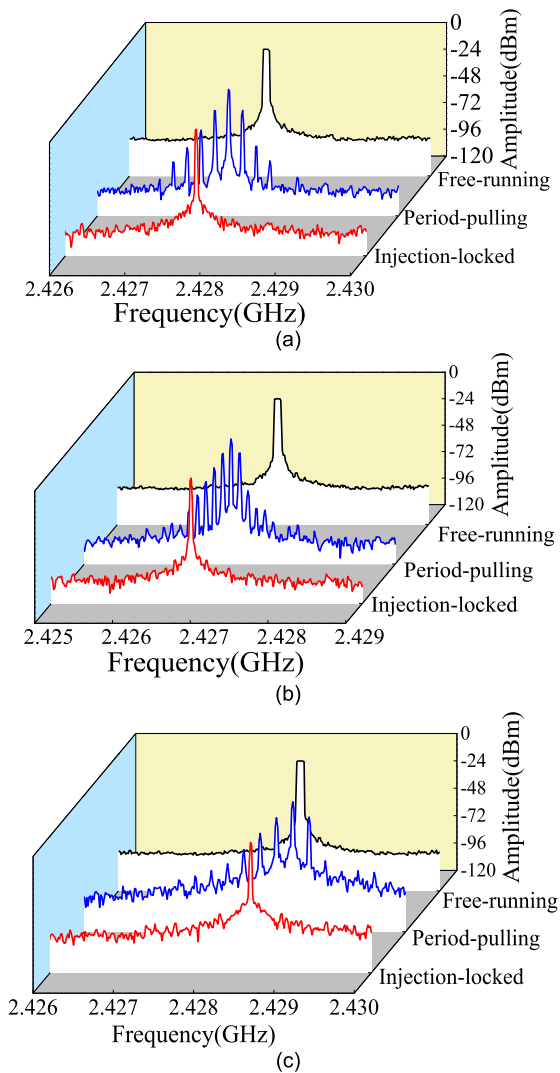

Fig. 11. Magnetron output spectra of different systems. (a) Conventional system. (b) System with monopole probe. (c) System with near-field coupling ring.

TABLE II
LOCKING BANDWIDTH OF DIFFERENT METHODS

| $\rho$ | Structure #1 | Structure #2 | Traditional |
| --- | --- | --- | --- |
| 0.06 | 0.12 MHz | 0.10 MHz | 1.14 MHz |
| 0.1 | 0.30 MHz | 0.17 MHz | 2.12 MHz |
| 0.2 | 0.70 MHz | 0.30 MHz | 3.92 MHz |

two structures. The monopole probe injects the microwave power by exciting certain resonant mode in the filter box, and coupling ring transmits the power to the receiving ring, which is connected with the filament leads. $|S_{11}|$ of the two different structures cannot be used as a parameter for comparing the performance.

The injection-locking experiment results are shown in Fig. 11. The frequency of injected signal is 2.4277, 2.4268, and 2.4285 GHz since the slight shifting of the magnetron free-running frequency in each experiment. We observed the spectrum of the magnetron output at different states (e.g., the free-running state, the periodic-pulling state, and the injection-locked state) with different injection systems and structures. The periodic-pulling state is an intermediate state between free running and injection locked; the sideband of the state contains frequency modulation and amplitude modulation effects as depicted in the spectrum. The output frequency of the magnetron is locked at the frequency of the injected signal. Regardless of the injection method and structure, the magnetron output spectrum is almost identical. The results indicate that the injection effects of the filament-injection method do not differ from those of the conventional one. Therefore, the proposed system and the two injection structures are practical and feasible.

Table II compares the locking bandwidth under three injection-locking experiments. The locking bandwidths of the two injection-locking structures are narrower than the conventional method. Referring to the analysis of the locking bandwidth in (13), the difference in the locking bandwidth occurs because of the coefficient $k$. For the same injection ratio, most of the injection energy in the proposed method is dissipated. The loss of the proposed two structures is different. For the monopole probe structure, the main source of the loss is return loss. About half of the injected power is reflected. Also, a part of power is dissipated in the filter box as heat loss. For the coupling ring structure, the return loss of it is small, but the injected power is also small. This is may be caused by the radiation loss, a part of injected power is radiated in the filter box. There is also some of the injected power transferred into heat because of the FR4 substrate. How to reduce attenuation of the injection structure is one of the targets in the future.

The locking bandwidth of the monopole probe structure is greater than that of the near-field coupling ring structure, consistent with the cold test results but not with the simulation. The phenomenon may be caused by the imperfect simulation of the leads impedance and the failure to fully calculate the resonant mode in the filter box, which is actually a multimode cavity. The current simulation method can only qualitatively but not quantitatively reflect the performance of a certain structure, necessitating further improvement.

The near-field coupling ring structure cannot work for an extended duration because the breakdown voltage of the FR4 will decrease over time. The substrate may be broken down after several minutes because the heat from the filament leads to a temperature rise of the PCB. Because there is a dc voltage of approximately 4000 V between the top and bottom sides of PCB, the experiments should be performed carefully. The problem may be solved in the future using a ceramic substrate with higher breakdown voltage and stable performance at higher temperatures.

The experiment confirms the feasibility of the proposed injection-locked magnetron system. The filament-injection method follows the same law as the conventional one. Currently, the injection-locking bandwidth is relatively narrow; the value of $k$ may be improved with a superior injection structure designed in the future. The locking bandwidth will be comparable to conventional injection methods. Table III presents the comparison of the proposed and conventional injection-locked magnetron systems. The weight of the proposed system is reduced by 8.4 kg, and the volume[2] of the system is reduced to 16% of the conventional system. Accordingly, the cost of the system is also significantly reduced because there is no waveguide circulator and waveguide-coaxial converter. The filament-injection magnetron system achieves the miniaturization and cost reduction of an injection-locking magnetron system.

---

[2]The volume is calculated by the maximum length, width, and height of the system.

TABLE III
COMPARISON OF TRADITIONAL AND PROPOSED INJECTION-LOCKED MAGNETRON SYSTEMS

|  | Waveguide Circulator | Waveguide-coaxial convertor | Weight(kg) | Volume(cm$^3$) |
|---|---|---|---|---|
| Traditional system | yes | yes | 12.1(Magnetron, excitation cavity, circulator, coaxial-waveguide adapter, dummy load) | 81.0×52.0×16.8 |
| Proposed system | no | no | 3.4(Magnetron, excitation cavity, dummy load) | 42.2×16.1×16.8 |

## V. CONCLUSION

A novel injection-locked magnetron system based on a novel injection method is proposed, and the system is developed and measured. To our knowledge, it is the first time that a magnetron filament and filter box have been used to inject reference signals. Two feasible structures are designed, fabricated, and measured. The maximum locking bandwidth of 0.7 MHz is achieved. The system is compact and economical compared to conventional systems.

To our knowledge, it is also the first injection-locked magnetron system with no waveguide isolation devices between the reference injection signal and the magnetron output microwave. The proposed injection-locking system reduces high-power microwave waveguide components, significantly decreasing the cost and size of the magnetron's injection-locking system, and provides a novel idea for developing an injection-locked magnetron.

In the future, magnetrons may be designed to have a high-efficiency injection-locking path from their filaments, and the injection-locking circuits may be integrated into the filter box. Also, the phase-locking characteristic of the system will be studied and experimented. The injection-locked magnetron is a candidate solution to solve high-cost and large-volume drawbacks of conventional injection-locked magnetron systems in microwave heating applications. Moreover, especially in microwave ovens, it only requires to enhance the filter boxes of commercial magnetrons without any change to the magnetron main component, e.g., the resonant cavity, but will efficiently improve the controllability of microwave heating through frequency-selective heating by means of seeking a frequency to less reflection.